\documentclass[reprint,superscriptaddress,amssymb,amsmath,aps,prx,longbibliography]{revtex4-2}
\usepackage{minitoc}

\usepackage{graphicx}
\usepackage{dcolumn}
\usepackage{bm}
\usepackage{siunitx}
\usepackage{booktabs}
\usepackage[usenames,dvipsnames]{xcolor}
\usepackage{tcolorbox}
\usepackage{tabularx}
\usepackage{array}
\usepackage{colortbl}
\usepackage{braket}
\usepackage{gensymb}
\usepackage{pdfpages}
\usepackage{amsmath}
\usepackage{mathrsfs}
\usepackage{blindtext}
\usepackage{minitoc}
\usepackage{dsfont}
\usepackage{multirow}
\usepackage{xcolor}
\usepackage{soul}
\usepackage[normalem]{ulem}

\usepackage{amssymb}

\makeatletter
\patchcmd{\@outputpage@head}{\@ifx{\LS@rot\@undefined}{}{\LS@rot}}{}{}{}
\makeatother

\tcbuselibrary{skins}

\newcommand{\units}[1]{\,\mathrm{#1}}
\tcbset{tab2/.style={colback=gray!5!white,colframe=black!50!black,colbacktitle=Gray!40!white,
coltitle=black,center title}}

\begin{document}

\title{Three-Wave Mixing Element with Quantum Paraelectric Materials}

\author{Eric I. Rosenthal}
\email{ericros@stanford.edu}
\affiliation{E. L. Ginzton Laboratory, Stanford University, Stanford, California 94305, USA}
\affiliation{Present address: Sygaldry Technologies, Ann Arbor MI, USA}

\author{Christopher S. Wang}
\affiliation{James Franck Institute and Department of Physics, University of Chicago, Chicago, Illinois, 60637, United States}
\affiliation{Present address: Q-NEXT, Argonne National Laboratory, Lemont, Illinois 60439, USA}

\author{Jamison Sloan}
\affiliation{E. L. Ginzton Laboratory, Stanford University, Stanford, California 94305, USA}

\author{Giovanni Scuri}
\affiliation{E. L. Ginzton Laboratory, Stanford University, Stanford, California 94305, USA}

\author{Yueheng Shi}
\affiliation{E. L. Ginzton Laboratory, Stanford University, Stanford, California 94305, USA}

\author{Kaveh Pezeshki}
\affiliation{E. L. Ginzton Laboratory, Stanford University, Stanford, California 94305, USA}

\author{Peter Mugaba Noertoft}
\affiliation{E. L. Ginzton Laboratory, Stanford University, Stanford, California 94305, USA}

\author{Jelena Vu\v{c}kovi\'{c}}
\affiliation{E. L. Ginzton Laboratory, Stanford University, Stanford, California 94305, USA}

\author{Christopher P. Anderson}
\email{cpand@illinois.edu}
\affiliation{Department of Materials Science and Engineering, University of Illinois Urbana-Champaign, Urbana, Illinois 61801, USA}

\date{\today}
\begin{abstract}
At cryogenic temperatures and microwave frequencies, the perovskite crystals strontium titanate (STO) and potassium tantalate (KTO) have large, tunable permittivity arising from a quantum paraelectric phase. As such, these materials hold promise as a platform to realize compact, variable capacitance elements for use as three-wave mixing elements in quantum devices. From modulating this capacitance within a resonant circuit, we propose the development of a quantum ``paraelectric nonlinear dielectric amplifier" (PANDA). We calculate that a PANDA made from a nanofabricated parallel plate capacitor and realistic design constraints can demonstrate a three-wave mixing strength of order MHz, in comparison to an effective Kerr strength of sub-Hz. This suggests excellent performance as a degenerate parametric amplifier, with high compression power in analogy to superconducting parametric amplifiers based on kinetic inductance. Beyond parametric amplifiers, we predict that compact, tunable capacitors based on STO, KTO, and related materials can enable a wide class of cryogenic quantum circuits including novel filters, switches, circulators, and qubits.
\end{abstract}
\maketitle

\section{Introduction}
The growth of quantum technology has created the need for cryogenic circuit elements that are compact, tunable, and low loss. Superconducting quantum processors, for example, are a leading platform for quantum information, but operate under a stringent set of requirements such as millikelvin temperatures, limited external fields, and signals at the level of single microwave photons \cite{kjaergaard:2020}. This motivates the development of devices engineered to have improved performance and scalability in these regimes.

Superconducting qubits are based on the Josephson junction (JJ), an inductive nonlinearity. JJs are also the building block for much of the circuitry adjacent to the qubits, including the quantum-limited parametric amplifiers. These amplifiers are required for qubit readout \cite{clerk:2010,aumentado:2020}, and, have utility for fundamental science, e.g. axion dark-matter detection \cite{malnou:2018}. As such, there is a growing body of literature on improving the gain, bandwidth, power handling, and scalability of parametric amplifiers, all while pushing noise performance as close as possible to the quantum limit \cite{macklin:2015,frattini:2018,eddins:2019,planat:2019,rosenthal:2021,lecocq:2020b,white:2023,wang:2025}.

However, JJ-based parametric amplifiers are not suitable for all applications. The compression power, marking the onset of nonlinear saturation, is typically limited to below approximately -100 dBm \cite{aumentado:2020}. JJs are also susceptible to the breakdowns in superconductivity that occur under high magnetic fields and optical illumination. Yet, there are many applications for quantum-limited measurement under these conditions. Microwave-to-optical transducers require close interplay between microwave circuits and optical fields \cite{mirhosseini:2020,delaney:2022}, and magnetic fields are essential to the study of solid-state spin defects \cite{wang:2023}, axion dark-matter haloscopes \cite{malnou:2019,backes:2021}, and Majorana qubits \cite{microsoft:2025}, for example.

The rich application space for quantum-limited parametric amplifiers motivates the development of alternative parametric mixing elements \cite{frattini:2017,frattini:2018}. Alternative nonlinearities to the Josephson junction include kinetic inductance \cite{malnou:2021,parker:2022,malnou:2022,xu:2023,frasca:2024,khalifa:2024,splitthoff:2024}, quantum dots \cite{cochrane:2022}, field-effect transistors \cite{phan:2023}, and hybrid superconductor-semiconductor junctions \cite{hao:2024}. In particular, kinetic inductance parametric amplifiers (KIPAs) have recently been shown to achieve high gain with low noise at compression powers of approximately -50 dBm and tesla-scale magnetic fields \cite{parker:2022,vaartjes:2024,frasca:2024}.

Parametrically tunable capacitors (varactors) are a conceivable alternative to these nonlinearities. Varactors have long appeared within classical, room-temperature electrical engineering literature \cite{blackwell:1961,penfield:1962}, including work from the mid-$\mathrm{20^{th}}$ century that makes use of strontium titanate $\mathrm{SrTiO_3}$ (STO) \cite{billeter:1964,vendik:1972}. Yet despite this storied history, varactors have thus far received limited attention within the context of quantum information experiments that operate at microwave frequencies and millikelvin temperatures, or that utilize quantum paraelectric phases.

Here, we propose using the tunable capacitance of quantum paraelectric materials as an alternative parametric mixing element for quantum circuits. In particular, we propose the development of quantum paraelectric nonlinear dielectric amplifiers (PANDAs) as an alternative parametric amplifier for use in quantum technologies. Building on Ref.~\cite{ulrich:2024}, we analyze the design of a PANDA realized from a tunable capacitor (varactor) made from the perovskites strontium titanate, $\mathrm{SrTiO_3}$ (STO) or potassium tantalate, $\mathrm{KTaO_3}$ (KTO), quantum paraelectric materials which have large voltage tunable dielectric permittivities. We derive the nonlinear Hamiltonian for a resonator made from an STO or KTO capacitor, and calculate that a nanofabricated capacitor geometry can display three-wave-mixing strengths (i.e. parametric coupling rate) as great as approximately $O(10 \, \textrm{MHz})$ with approximately $O(0.1)\units{Hz}$ effective Kerr terms (a ratio of $10^{8}$). This is greater nonlinearity than varactors made from bulk substrates \cite{apostolidis:2024}, and suggests high parametric gain and competitive dynamic range compared to state-of-the-art KIPAs \cite{parker:2022,vaartjes:2024,frasca:2024}. Overall, our proposal examines the rich design space of rf/microwave circuits made from perovskite dielectrics and related materials, motivates open challenges in the nanofabrication of these materials, and serves as a blueprint for future experiments that seek to realize these emerging cryogenic components.

\begin{figure}[htb!] 
\begin{center}
\includegraphics[width=1.0\columnwidth]{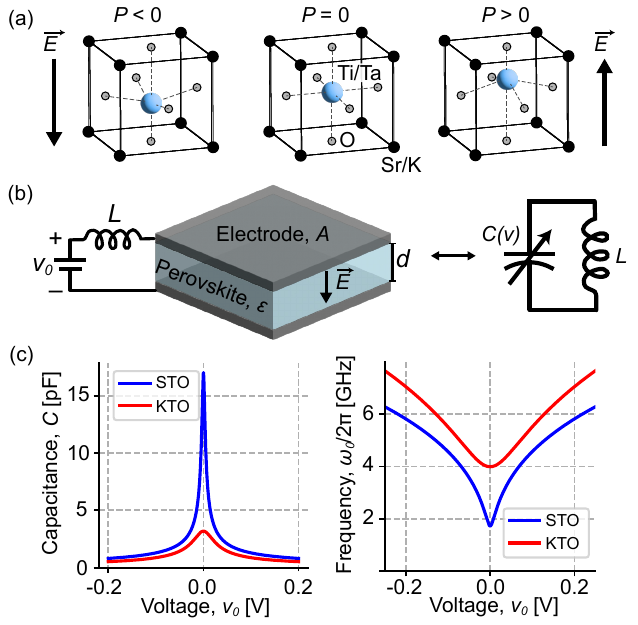}
\caption{
(a) The perovskite crystals strontium titanate (STO) and potassium tantalate (KTO) have a cagelike structure where a central Ti or Ta atom lies within a cell formed by O and Sr or K atoms, respectively. At cryogenic temperatures these materials have high permittivity and low loss. Permittivity is tunable via an applied electric field $\Vec{E}$, which displaces the central atom and polarizes the crystal. (b) A voltage tunable resonator is formed by a parallel plate capacitor made from thin-film STO or thin-film KTO, and, a coupling inductance $L$. An external voltage bias $v_0$ across the plates induces an electric field $E=v_0/d$ and tunes the capacitance. (c) Capacitance and resonance frequency $\omega_0=1/\sqrt{LC(v)}$ as a function of $v_0$ for $A=(4 \units{\mu m})^2$ and $d=200 \units{nm}$ with the dielectric properties given in Appendix~\ref{sec:dielectric_properties}. Parameters are chosen to enable microwave frequency operation and a compact footprint.
}
\label{fig:resonator}
\end{center}
\end{figure}

\section{Dielectric Properties}
STO and KTO have a similar structure in which a Ti/Ta atom is centered within an octahedral cage of O atoms, Fig.~\ref{fig:resonator}a. In the paraelectric phase, the material becomes polarized when the central atom is displaced by an electric field $E$, changing the dielectric properties. The resulting normalized bias is parameterized by $\lambda=\sqrt{\lambda_s^2 + (E/E_N)^2} \approx E/E_N$, where $E_N$ is the re-normalizing field and $\lambda_s$ is a dimensionless measure of material inhomogeneity, that goes to zero for an ideal crystal. Permittivity and loss depend on bias as described by a well-known modified Landau-Ginzburg-Devonshire (LGD) theory \cite{vendik:1997,vendik:1997b,vendik:1999,geyer:2005,fujishita:2016}, with permittivity approximated by:
\begin{multline}
\varepsilon(\lambda)/\varepsilon_{00} = \\ \left(\left(\sqrt{\lambda^2+\eta^3}+\lambda \right)^{2/3} + \left(\sqrt{\lambda^2+\eta^3}-\lambda \right)^{2/3} - \eta\right)^{-1},
\label{eq:epsilon}
\end{multline}
where $\varepsilon_{00}$ and $\eta$ are material parameters. See Appendix~\ref{sec:dielectric_properties} for details and a related model for loss.
    
Because here we are interested in rf and microwave devices engineered to operate at cryogenic temperatures for quantum applications, we consider the LGD theory in the limit of low frequency (order 100 GHz and below \cite{neville:1972,kozina:2019}) and low temperature, only. From measurement of bulk crystals at 4 kelvin and at zero field ($\lambda=0$) these models \cite{geyer:2005,fujishita:2016,davidovikj:2017} predict a relative permittivity and loss tangent of $\varepsilon_r = \varepsilon/\varepsilon_0 \approx 24\times10^3$ and $\tan(\delta)\approx10^{-3}$ for STO, and $\varepsilon_r \approx 4.5\times10^3$ and $\tan(\delta)\approx10^{-5}$ for KTO. In the limit of large field ($\lambda\gg0$), $\varepsilon_r$ decreases to order one thousand and modifies the loss(Appendix~\ref{sec:dielectric_properties}). In reality, both permittivity and loss can depend on material properties including defect concentration and strain. Minimizing loss in both bulk and thin films is an open research challenge \cite{engl:2019,ulrich:2025,anderson:2025}. For example, Ref.~\cite{brahim:2025} reports permittivity of approximately $2\times10^3$ and a loss tangent of $3.4\times10^{-4}$ in thin-film STO at microwave frequencies.

\section{A Nonlinear Resonator} We propose to make use of these dielectric properties to develop compact and tunable capacitors made from STO, KTO, or related materials, and which can be integrated into resonant circuits. In this work, we consider a lumped-element resonator implemented with a parallel-plate capacitor (PPC) with area $A$ and an STO or KTO dielectric thin film of thickness $d$ with superconducting metallization on the faces (Fig.~\ref{fig:resonator}b). However our analysis is also applicable to a coplanar waveguide (CPW) design on a bulk STO or KTO substrate \cite{davidovikj:2017}, which differs only in the geometric design space. Moreover, we envision a galvanic connection to one of the metal faces, which enables electric-field tuning of the material with an external bias voltage.

\begin{figure}[htb!] 
\begin{center}
\includegraphics[width=1.0\columnwidth]{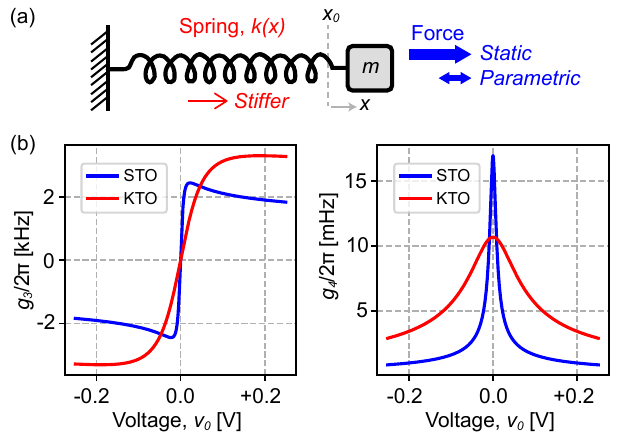}
\caption{(a) Our nonlinear resonant circuit is analogous to an oscillating mass on a spring whose spring constant stiffens when stretched. We are interested in its dynamics, Eq.~\ref{eq:Hamiltonian_taylor}, given both a static restoring force set by $v_0$, and, a parametric driving force set by $v_\mathrm{ac}$. (b) By Taylor expanding the potential around a variable bias $v_0$, we determine the nonlinear coefficients $g_3$ and $g_4$.}
\label{fig:nonlinearity}
\end{center}
\end{figure}

With an eye towards quantum-limited parametric amplification at approximately mK temperatures, we begin by highlighting our circuit design criteria. We target resonance frequencies in the approximately GHz range using the STO or KTO capacitance $C(v)=\varepsilon_0 \varepsilon_r(v) A / d$ (Fig.~\ref{fig:resonator}c). This capacitance must be widely tunable to enable commensurate nonlinear mixing capabilities. We neglect any stray capacitance to ground given that $\varepsilon_r \gg \varepsilon_\mathrm{substrate}$ for conventional substrate hosts (i.e. silicon or sapphire). To tune $C(v)$, it is convenient to work with volt-scale biases compatible with standard low-noise control electronics, which in turn influences the choice of dielectric thickness $d$. As such, varactors made from bulk material are not optimal for our applications ~\cite{apostolidis:2024}. We also limit film thicknesses to $>$100 nm, which have been shown to display bulklike low-temperature dielectric properties \cite{yang:2022}; thinner films may suffer from dead-layer effects. Finally, design parameters should be compatible with a reasonable fabrication process. 

Given these considerations, we choose as an illustrative example an inductance of $L$ = 0.5 nH, and capacitor dimensions of $A=(4 \units{\mu m})^2$ and $d=200 \units{nm}$ (a released film thickness of 105 nm was recently experimentally demonstrated for STO \cite{ulrich:2025}). This corresponds to a tunable capacitance $C(v)=\partial q/\partial v = A \varepsilon_0 \varepsilon_r(v)/d$ between $17.2$ and $1.28 \units{pF}$ for STO, and between $3.18$ and $0.86 \units{pF}$ for KTO using a bias range $\{0, 250\}$ mV and the dielectric model in Appendix~\ref{sec:dielectric_properties}. The magnitude of this tunability is comparable to that of inductive elements based on superconducting quantum interference devices (SQUIDs).

The resulting tunable, nonlinear resonator can be described by the following Hamiltonian:
\begin{align}
    H/\hbar = \omega_0a^\dagger a +ig_3\big(a - a^\dagger\big)^3 + g_4\big(a - a^\dagger\big)^4 + ...,
    \label{eq:Hamiltonian_taylor}
\end{align}
where $\omega_0$ is the resonator frequency, $g_3$ and $g_4$ are third- and fourth- order nonlinearities of interest, and $a$ ($a^\dagger$) are complex variables that can be cast in terms of the bosonic annihilation (creation) operators in the quantum mechanical description (see Appendix~\ref{sec:circuit_hamiltonian}). This Hamiltonian has strong analogy to the inductive mixing elements conventionally used in circuit quantum electrodynamics (QED), e.g. the SNAIL-based circuits in Ref.~\cite{frattini:2018}. The sign of the nonlinear terms in Eq.~\ref{eq:Hamiltonian_taylor} is based on the standard definition of the quadrature operators \cite{vool:2017}, and leads to a difference from conventional circuit QED devices because in this case the nonlinearity is capacitive rather than inductive. The parameters $\{\omega_0, g_3, g_4\}$ all depend on the PPC capacitance, and therefore can be tuned with $v_0$. As shown in Fig.~\ref{fig:nonlinearity}b, our proposed design can simultaneously achieve a high ratio between the third- and fourth-order nonlinearities, $g_3/2\pi \sim O(\textrm{kHz})$ and $g_4/2\pi \sim O(\textrm{mHz})$, and holds promise as a compact and high-performance three-wave mixing element for microwave frequency quantum circuits.

\section{A Degenerate Parametric Amplifier}
\begin{figure}[htb!] 
\begin{center}
\includegraphics[width=1.0\columnwidth]{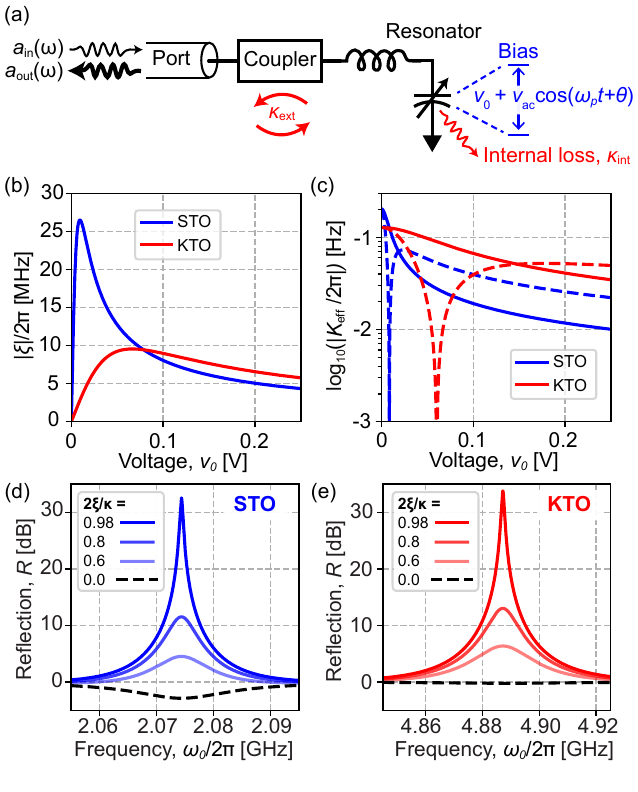}
\caption{
(a) Input-output model for a parametric amplifier made from an STO or KTO capacitor. Modulation of the capacitance around $\omega_p \approx 2\omega_0$ yields degenerate parametric amplification. (b) Magnitude of the three-wave-mixing (3WM) strength $|\xi|/2\pi$ choosing $v_\mathrm{ac} = 1\units{mV}$ (ac bias charge $q_\mathrm{ac}=v_\mathrm{ac}C(v)$). (c) Magnitude of the effective Kerr strength $K_\mathrm{eff}/2\pi$ using a rotating-wave approximation (solid) and one that includes second-order corrections due to $g_3$ (dashed). (d,e) Reflection $R(\omega)$ of an STO and KTO device, respectively, plotted for different values of $\xi$ normalized to resonator loss $\kappa/2$. The model (Eq.~\ref{eq:gain_3WM}) uses the parameters as given in Fig.~\ref{fig:nonlinearity} and an external quality factor of $Q_\mathrm{ext}=100$.
}
\label{fig:gain}
\end{center}
\end{figure}

Parametric amplification is a key potential use case for this nonlinear resonator (the PANDA). To use our circuit as a degenerate parametric amplifier, we will modulate the charge on the capacitor near twice the resonant frequency $\omega_p \sim 2\omega_0$ and with phase $\theta$ in the presence of an offset bias $v_{0}$. For simplicity, we consider a single-port implementation of our amplifier (Fig.~\ref{fig:gain}a). We can model the resulting dynamics of small perturbations around the charge equilibrium set by $v_0$ via
\begin{align}
    & H_\mathrm{driven}/\hbar = \omega_0 a^{\dagger}a + \frac{\xi}{2} a^{\dagger 2} + \frac{\xi^*}{2} a^{2} + \frac{K_\mathrm{eff}}{2} a^{\dagger 2}a^2, 
    \label{eq:Hamiltonian_nonzero_bias}
\end{align}
where $\xi=3 g_3 (v_\mathrm{ac}/v_\mathrm{zpf})e^{-i\theta}$ is the three-wave-mixing (3WM) strength, and $K_\mathrm{eff}=12 (g_4 - 5 g_3^2 / \omega_0)$ is the effective Kerr term (Appendix~\ref{sec:circuit_hamiltonian}). Here $v_\mathrm{ac}$ is the ac modulation voltage normalized by the resonator's zero-point voltage fluctuations $v_\mathrm{zpf}$, and $e^{-i\theta}$ arises from the phase of the pump. In Fig.~\ref{fig:gain} we plot these terms against the external dc bias $v_0$.

As is standard in the study of parametric amplifiers \cite{eichler:2014,planat:2019,parker:2022}, Eq.~\ref{eq:Hamiltonian_nonzero_bias}, is derived (Appendix~\ref{sec:circuit_hamiltonian}) within the rotating-wave approximation, and, at small enough amplitudes so that higher-order dynamics can be neglected. In our case, however, an expanded expression for $K_\mathrm{eff}$, which incorporates second-order corrections from the third-order nonlinearity $g_3$, goes to zero at a particular bias point (dashed line in Fig.~\ref{fig:gain}c) \cite{frattini:2021}. This suggests that there exists a range of external bias $v_0$ where the Kerr term can, in principle, be nulled while retaining a significant $\xi$. This cancellation, however, does not reflect an underlying symmetry-breaking structure to our circuit, such as in SNAIL devices \cite{frattini:2017,frattini:2018}. In Appendix~\ref{sec:design_considerations} we explore how modifying the circuit geometry impacts the quantities in Eq.~\ref{eq:Hamiltonian_nonzero_bias}.

The calculations shown in Fig.~\ref{fig:gain} indicate that a properly engineered and operated PANDA can be an excellent amplifier with minimal unwanted four-wave-mixing (Kerr) terms that degrades performance. Using the capacitor parameters given in Fig.~\ref{fig:resonator}c, and, as an illustrative example, an ac bias of $v_\mathrm{ac}=1~\units{mV}$, we find a maximum three-wave-mixing (3WM) strength of approximately 26 MHz for STO and 9.5 MHz for KTO. Our choice of $v_\mathrm{ac}$ translates to pumping with approximately $10^4$ photons for both materials; a precise comparison will depend on broader design considerations. Pumping with greater amplitude will increase the 3WM strength, which to the lowest order is proportional to $v_\mathrm{ac}$ (Appendix~\ref{sec:circuit_hamiltonian}). Too large a pump field, however, will eventually drive higher-order nonlinearities in a manner dependent on higher-order corrections to Eq.~\ref{eq:Hamiltonian_nonzero_bias}, which in general are dependent on dc bias point.

For this design, the optimal 3WM points occur at a dc bias of 9.3 mV for STO and 66 mV for KTO, respectively; modest voltages compatible with superconducting electronics. At these operation points $K_\mathbf{eff}$ is of order 0.1 Hz or less for both materials (Fig.~\ref{fig:gain}. For comparison, the Josephson parametric amplifier (JPA) in Ref.~\cite{rosenthal:2021}, operated as a linear amplifier, has $\xi/2\pi\lesssim 25 \units{MHz}$ and $K_\mathbf{eff}/2\pi\approx50\units{kHz}$, and the KIPA in Ref.~\cite{parker:2022} operates at $\xi/2\pi \lesssim 27 \units{MHz}$ and $K_\mathbf{eff}\approx0.1\units{Hz}$. Our maximum predicted figure-of-merit ratio of $\xi/K_\mathrm{eff} \approx 10^{8}$ is much higher than the $\xi/K_\mathrm{eff} \lesssim 10^5$ ratio typical of Josephson parametric amplifiers (JPA) \cite{boutin:2017,frattini:2018,planat:2019} and is comparable to state-of-the-art KIPAs \cite{malnou:2021,parker:2022,vaartjes:2024,frasca:2024}.

To analyze our circuit as a degenerate parametric amplifier, we consider the reflection of a weak signal tone while the capacitor is charge pumped as previously discussed. Following an input-output formalism applied to $H_\mathrm{driven}$ (Eq.~\ref{eq:Hamiltonian_nonzero_bias}), the circuit reflection coefficient is ~\cite{gardiner:1985,parker:2022},
\begin{equation}
    R(\omega) = \frac{\kappa_\mathrm{ext}\kappa/2+i\kappa_\mathrm{ext}(\Delta+\omega-\omega_p/2)}{\Delta^2+(\kappa/2+i(\omega-\omega_p/2))^2-|\xi|^2} -1
    \label{eq:gain_3WM}.
\end{equation}
Here $\Delta=\omega_0 - \omega_p/2$ is the detuning of the half-pump frequency from the resonator, and for the remaining analysis we choose $\Delta=0$.

A critical design parameter is the external coupling rate $\kappa_\mathrm{ext}$, which should be much larger than the internal loss rate $\kappa_\mathrm{int}$ for optimal noise performance. $\kappa_\mathrm{ext}$ will be determined by the details of the coupler, whereas $\kappa_\mathrm{int}$ depends on the material of choice. Our requirement of introducing both dc and ac bias fields across the capacitor translates to having an inductive coupler. For example, $\kappa_\mathrm{ext}$ can be controlled by implementing the coupler as a stepped-impedance filter (see Fig.~\ref{fig:layout}), in further analogy to kinetic inductance amplifiers. Advantageously, this bias scheme avoids challenges with magnetic flux control (including nonuniformity of the bias field and crosstalk between flux lines) that constrain JPAs, e.g. as in Ref.~\cite{malnou:2019}. For the estimates in this paper, we consider an external quality factor $Q_\mathrm{ext}=\omega_0/\kappa_\mathrm{ext}=100$ for both designs, as is typical for superconducting parametric amplifiers \cite{planat:2019,parker:2022,vaartjes:2024,frasca:2024}. 

We expect internal loss to be dominated by bulk dielectric loss. The internal quality factor $Q_\mathrm{int}=\omega_0/\kappa_\mathrm{int}=\tan^{-1}{(\delta)}$ is evaluated at $v_0$ to be $Q_\mathrm{int}\approx6.1\times10^{2}$ for STO and $Q_\mathrm{int}\approx7.4\times10^{3}$ for KTO (from the model plotted in Fig.~\ref{fig:dielectric}c). These parameters yield $\kappa_\mathrm{int}/2\pi=3.4\units{MHz}$ and $\kappa_\mathrm{ext}/2\pi=20.7\units{MHz}$ for the STO design, and $\kappa_\mathrm{int}/2\pi=0.7\units{MHz}$ and $\kappa_\mathrm{ext}/2\pi=48.9\units{MHz}$ for the KTO design. (See Fig.~\ref{fig:kappa_vs_freq} for a model of internal loss rate versus resonator frequency.) Both designs are over-coupled ($\kappa_\mathrm{ext}/\kappa_\mathrm{int}>1$), but the KTO design is more overcoupled due to its lower modeled internal loss. This suggests that PANDAs made from KTO may have better noise performance, as internal loss scales added noise by a factor of $(\kappa_\mathrm{int} + \kappa_\mathrm{ext})/\kappa_\mathrm{ext}$ assuming both the external port and internal bath modes are thermalized to the same temperature \cite{eichler:2014}.

In Figs.~\ref{fig:gain}d,e we plot $R(\omega)$ for both STO and KTO capacitors, choosing the dc bias voltage $v_0=v_\mathrm{0,max}$ to maximize $\xi$ for each design. For this operating point we simulate a center frequency and 3-dB bandwidth of $\omega_0/2\pi=2.072 \units{GHz}$ and $\kappa/2\pi=24 \units{MHz}$ for the STO design, respectively, and $\omega_0/2\pi=4.882 \units{GHz}$ and $\kappa/2\pi=49.5 \units{MHz}$ for the KTO design, respectively. The operating frequency can be tuned \textit{in situ} by changing the dc bias. It can also be modified by designing different capacitor dimensions or a different inductance.

The compression power of a degenerate parametric amplifier is a critical parameter that determines the maximum gain and operating range of the amplifier. It is set by a combination of effects including Kerr nonlinearity, higher-order terms in the circuit Hamiltonian, and pump depletion; as such it is complicated to precisely model. Inductive nonlinearities, e.g. kinetic inductance of the electrodes and the proposed stepped-impedance coupler, may also affect power compression. Though we present the existence of a Kerr-free operating point, we estimate from lowest-order theory, only, that a PANDA can have an effective Kerr term of order $K_\mathrm{eff}/2\pi \approx 0.1 \units{Hz}$, in comparison to a linewidth of $\kappa/2\pi \approx 20 \units{MHz}$ as shown in Fig.~\ref{fig:gain}. It will therefore require $N\approx10^{8}$ circulating photons to shift the resonator frequency by a linewidth. At a resonant frequency of 2 GHz, this corresponds to a circulating power of $P_\mathrm{circ} = N \hbar \omega_0 \kappa \approx -64 \units{dBm}$, similar to state-of-the-art KIPAs (e.g. 1-dB compression powers of -50 dBm in Ref.~\cite{parker:2022}, -65 dBm in Ref.~\cite{frasca:2024}), and greater than JJ-based parametric amplifiers engineered for high-power compression (e.g. approximately -100 dBm in Ref.~\cite{frattini:2018}). A three-wave-mixing element based on a nanoscale capacitor made from STO or KTO can therefore be expected to have power handling competitive with respect to superconducting alternatives, along with advantages from a compact form factor. To optimize power handling it is helpful to use thick electrodes to reduce kinetic inductance, and to use high-quality materials with large dielectric susceptibility and low loss.

\section{Conclusion}
We propose the development of a parametric mixing element based on the tunable permittivity of the quantum paraelectric perovskite materials STO and KTO. First, we derive an interaction Hamiltonian for a novel parametric device, realized by integrating this nonlinear element into a resonant circuit. We calculate voltage-tunable third- and fourth-order nonlinear terms that allow for three-wave- or four-wave-mixing dynamics, respectively, in analogy with superconducting parametric amplifiers based on the Josephson effect or kinetic inductance. We predict that the third-order (fourth-order) terms can have order MHz (sub-Hz) magnitude, a large ratio which suggests promise as a degenerate parametric amplifier (PANDA) with favorable dynamic range compared to superconducting alternatives. Due to the high permittivity of STO and KTO, a PANDA can also have a compact layout even at sub-GHz frequencies. We also note that with related materials or modifications from isotope exchange, strain, doping can lead to even stronger nonlinearities and lower loss near zero temperature than modeled here \cite{anderson:2025}.

While we focus specifically on a simple parametric amplifier circuit, we expect the large and tunable permittivity of STO and KTO to be widely useful in other rf and microwave frequency quantum circuits including filters, switches and modulators, nonreciprocal devices, and qubits. Unlike superconducting nonlinearities, STO and KTO are, in principle, robust to optical illumination and magnetic fields \cite{tinsman:2016,yang:2025}; PANDAs can thus be of use for applications such as microwave-to-optical transduction or axion dark-matter detection, where robust parametric mixing elements are desired. We predict application to next-generation axion dark-matter haloscopes will be especially fruitful; the combined need for tesla-scale magnetic fields, quantum-limited microwave measurement, and microwave squeezed light in modern experiments \cite{backes:2021}, remains a challenging operational limit for conventional superconducting parametric amplifiers. In summary, our proposal highlights the need for materials and nanofabrication development of STO and KTO microwave devices, and the unexplored application of these devices toward compact and scalable quantum technologies.

\vspace{0.1in}
\section{Acknowledgments}
The work in the Vu\v{c}kovi\'{c} group has been supported by the Department of Energy under the Q-NEXT program and by a Google AI Research grant. C.S.W. acknowledges the support of the Grainger Fellowship from the University of Chicago. G.S. acknowledges support from the Stanford Bloch Postdoctoral Fellowship. C.P.A is supported by AFOSR FA9550-25-1-0309. The views expressed in the article do not necessarily represent the views of the U.S. DOE or the United States Government. We thank Christian Haffner, Nathaniel Kinsey, Andrew Higginbotham, Osmond Wen and Nicholas Frattini for helpful discussions.

\appendix
\label{appendix}

\section{Dielectric Properties}
\label{sec:dielectric_properties}
\begin{figure}[htb! ] 
\begin{center}
\includegraphics[width=1.0\columnwidth]{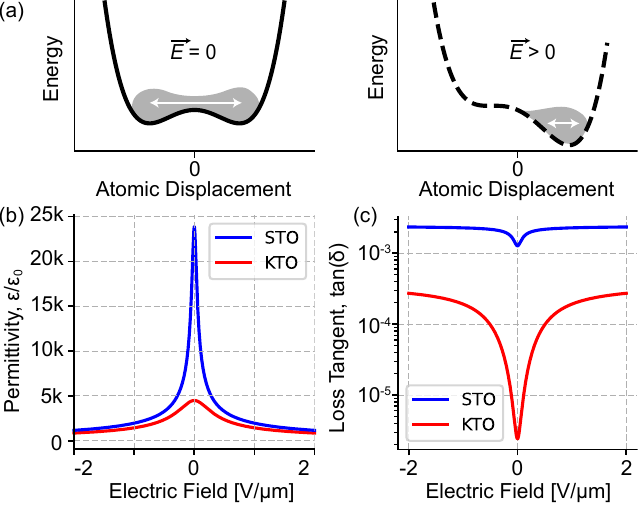}
\caption{At cryogenic temperature the perovskites STO and KTO have multiple low-energy states associated with positive or negative displacement of the central atom \cite{geyer:2005,fujishita:2016}. This leads to a quantum paraelectric phase in which the quantum fluctuations (horizontal white arrows) are significant compared to the energy barrier between these two potential wells. Permittivity is proportional to the inverse curvature of the free energy compared to relative displacement. An electric field, $E$, tilts the potential, which constraints the magnitude of fluctuations and reduces permittivity. STO and KTO differ in the depth of the double-well potential and its susceptibility to an applied field~\cite{esswein:2022}. (b) Relative permittivity $\varepsilon_r=\varepsilon/\varepsilon_0$ and (c) loss tangent $\mathrm{tan}(\delta)$, modeled vs $E$. Calculations use Eq.~\ref{eq:epsilon_vs_bias}, which is based on the model and material parameters in Ref.~\cite{vendik:1999}.
}
\label{fig:dielectric}
\end{center}
\end{figure}

The dielectric properties STO and KTO are well described by a modified Landau-Ginzburg-Devonshire (LGD) theory \cite{fujishita:2016}. In this model, the permittivity is inversely proportional to the curvature of the Gibbs free energy as a function of atomic displacement \cite{damjanovic:1998,esswein:2022}. At room temperature, both STO and KTO are in a paraelectric phase, where the net atomic displacement is zero. Both materials approach a phase transition at cryogenic temperatures. However, this transition is suppressed by quantum fluctuations, resulting a quantum paraelectric state. In these systems, the free energy is described by two potential wells separated by a barrier, and which correspond to two directions of atomic displacement, Fig.~\ref{fig:dielectric}a. At low temperatures, the energy barrier is weak enough compared to quantum fluctuations such that the atom occupies a superposition of both wells \cite{vendik:1997,vendik:1997b}. 

Following work and notation by Vendik \textit{et al.} \cite{vendik:1999b}, the dielectric constant of both STO and KTO are modeled using the modified LGD theory as
\begin{equation}
    \varepsilon(\lambda) = \frac{\varepsilon_{00}}{G(\lambda)^{-1} + i \Gamma(\lambda)} 
    \label{eq:epsilon_vs_bias}
\end{equation}
where $\varepsilon_{00}$ is a constant, $\Gamma(\lambda)$ describes dielectric losses and is small in magnitude compared to $G(\lambda)$, which is the real part of the Green's function defining the dielectric response:
\begin{multline}
    G(\lambda) = \\
    \left(\left(\sqrt{\lambda^2+\eta^3}+\lambda \right)^{2/3} + \left(\sqrt{\lambda^2+\eta^3}-\lambda \right)^{2/3} - \eta\right)^{-1}.
    \label{eq:G_vs_bias}
\end{multline}
Since the imaginary component of $\epsilon(\lambda)$ is small compared to the real term, we approximate permittivity as $\epsilon(\lambda)\approx\varepsilon_{00} G(\lambda)$ in Eq.~\ref{eq:epsilon} of the main text. Here $\eta\approx(\theta_F/T_c)\sqrt{1/16 + (T/\theta_F)^2}-1$, when $T<<\theta_F$, where $T$ is temperature, $\theta_F$ is the Debye temperature, and $T_c$ is the Curie temperature. For all models here, we assume $T\to0$ and as such, $\eta\approx\theta_F/(4T_c)-1$. The variable
\begin{equation}
    \lambda=\sqrt{\lambda_S^2 + (E/E_N)^2},
    \label{eq:normalized_bias}
\end{equation}
describes the normalized biasing from the unpolarized state, where $E$ is the bias field, $E_N$ is a renormalizing field, and $\lambda_S$ is a function of defects/inhomogeneity. For a pure material with negligible defect density, then $\lambda_S\approx0$ and $\lambda \to E/E_N$. Finally, the predicted frequency rolloff of the large dielectric response relates to the soft phonon frequency approxiately equal to THz, so at the relevant frequencies in the GHz range this assumption holds. 

According to Ref.~\cite{vendik:1999}, the loss term $\Gamma(\lambda)=\Gamma_1(\lambda) + \Gamma_2(\lambda) + \Gamma_3$ in Eq.~\ref{eq:epsilon_vs_bias} is a combination of the following. (1) Loss related to multiphonon scattering, which scales with temperature such that $\Gamma_1(\lambda)=A_1 (T/T_c)^2 G(\lambda)^{1/2}$, where $A_1$ is a constant. (2) Residual piezoelectricity, such that $\Gamma_2(\lambda) = A_2 y(\lambda)$ where $A_2$ is a constant and
\begin{multline}
    y(\lambda) =
    \left(\sqrt{\lambda^2+\eta^3}+\lambda \right)^{1/3} - \left(\sqrt{\lambda^2+\eta^3}-\lambda \right)^{1/3}
    \label{eq:y_vs_bias}
\end{multline}
is a residual ferroelectric displacement. Finally, (3) piezoelectricity due to charged defects such that $\Gamma_3 = A_3 n_d$, where $n_d$ is the density of charged defects and $A_3$ is another proportionality constant. The overall loss tangent is therefore modeled as
\begin{align}
    \mathrm{tan}(\delta) &= \mathrm{tan}(\delta_1) + \mathrm{tan}(\delta_2) + \mathrm{tan}(\delta_3),
    \label{eq:loss_tangents}
\end{align}
with the following terms,
\begin{align*}
    \quad \mathrm{tan}(\delta_1) &= A_1 (T/T_c)^2 G(\lambda)^{3/2}, \\
    \mathrm{tan}(\delta_2) &= A_2 y(\lambda)^2 G(\lambda), \\
    \mathrm{tan}(\delta_3) &= A_3 n_d G(\lambda).
\end{align*}

In Fig.~\ref{fig:dielectric} we plot models of the permittivity and loss tangent as a function of biasing electric field, $E$, based on Eq.~\ref{eq:epsilon_vs_bias}, and assuming operation at approximately $10 \units{mK}$ in the base of a dilution refrigerator. This model is based on Ref.~\cite{vendik:1999}, whose parameters are summarized in Table~\ref{tab:parameters}. From this model, we can see that both STO and KTO at millikelvin temperatures can have very high permittivity (e.g. $\varepsilon_r \approx 25\times10^{3}$ for STO and $\varepsilon_r \approx 5\times10^{3}$ for KTO), compared to the conventional dielectrics used in quantum devices (e.g. silicon or sapphire, which have $\varepsilon_r\approx10$). This permittivity is widely tunable with an applied electric field, and, especially for KTO, can be relatively low loss with $\delta \approx 10^{-5}$ \cite{geyer:2005}. 

These dielectric properties are relevant at rf and microwave frequency regimes and lower (e.g. characterized at 50 MHz in Ref.~\cite{neville:1972}, 1 GHz in Ref.~\cite{engl:2019}, 3 GHz in Ref.~\cite{vallabhapurapu:2021}, and between 4 and 6 GHz in Ref.~\cite{brahim:2025}). At frequencies of order THz and higher, the permittivity decreases due to oscillations associated with optical phonon modes in the crystal, and at optical frequencies STO has a refractive index of only 2.4 \cite{anderson:2025,ulrich:2025}. To the author's knowledge, the precise transition between these two limits has not yet been experimentally characterized but is predicted to be above 100's of GHz \cite{neville:1972,kozina:2019}.

\begin{table}[htb!]
\caption{Relevant parameters for STO and KTO, summarized from the model by Vendik \textit{et al} in Ref.~\cite{vendik:1999}. We assume the ideal case of no charged defects such that $n_d=0$. In general, these parameters may be highly dependent on material considerations including defect concentration, type, and strain. These parameters serve as an input to models in this paper, and inform the materials and fabrication considerations needed to optimize the proposed device performance.}
  \begin{center}
    \begin{tabular}{ |p{3.2cm}|p{1.2cm}|p{1.8cm}|p{1.8cm}|}
         \hline
         Description & Symbol & $\mathrm{SrTiO_3}$ & $\mathrm{KTaO_3}$ \\
         \hline
         Free parameter & $\varepsilon_{00}$ & $2080 \times \varepsilon_0$ & $1390 \times \varepsilon_0$ \\
         \hline
          Curie temp. [K] & $T_c$ & 42 & 32.5 \\
         \hline      
          Debye temp. [K] & $\theta_F$ & 175 & 170 \\
         \hline
          Renorm. field [V/$\mu$m] & $E_N$ & 1.93 & 1.56 \\
         \hline 
          Inhomogeneity & $\lambda_s$ & 0.018 & 0.020 \\
         \hline          
          Free parameter & $A_1$ & $2.45\times10^{-4}$ & $2.06\times10^{-4}$ \\
         \hline 
          Free parameter & $A_2$ & $2.45\times10^{-3}$ & $4\times10^{-4}$ \\
          \hline 
          Free parameter & $A_3$ & N/A & N/A \\
          \hline 
          Charged defects & $n_d$ & 0 & 0 \\
         \hline
          Operating temp. [K] & $T$ & $10^{-2}$ & $10^{-2}$ \\
         \hline         
    \end{tabular}
  \label{tab:parameters}
  \end{center}
\end{table}

\section{Derivation of Charge vs. Voltage Relation}
\label{sec:appendix_charge_vs_voltage}

We consider an idealized parallel plate capacitor that contains a dielectric characterized by a field-dependent permittivity $\varepsilon(E) = \varepsilon_0 \varepsilon_r(E)$ and a general nonlinear susceptibility $\chi(E) = \varepsilon_r(E)-1$. This dielectric has a strongly nonlinear polarization response,
\begin{equation}
    P(E) = \varepsilon_0 \int_0^{E} \chi(E') dE',
\end{equation}
noting that $\partial P / \partial E = \varepsilon_0 \chi(E)$. 

The electric and displacement fields are related as:
\begin{equation}
    D = \varepsilon_0 E + P(E),
    \label{eq:displacement_field}
\end{equation}
The divergence of $D$ is equal to the free charge density: $\nabla \cdot D = \rho_{\text{free}}$. The only free charges will be found on the surface, so using the integral form of Gauss's law over one of the plate capacitors gives the relationship:
\begin{equation}
    q = DA,
    \label{eq:charge_vs_displacement}
\end{equation}
where $q$ is the total charge on the plate, and $A$ is the plate area. Hence, we see that $D$ inside the capacitor is constant (a known result). Substituting Eq.~\ref{eq:displacement_field} into Eq.~\ref{eq:charge_vs_displacement} yields: $q(E) = A \varepsilon_0 \int_0^{E} \varepsilon_r(E') dE'$. Since $D$ is constant between the plates, and the material fills the entire capacitor, $E$ must also be uniform between the plates. Thus, the uniform electric field in the material can be related to the voltage $v$ between the plates as
\begin{equation}
    E = v/d.
\end{equation}
Combining the above relationships lets us write the charge-voltage relationship of the capacitor in terms of geometric parameters and the dielectric constant:
\begin{equation}
    q(v) = \frac{A \varepsilon_0}{d} \int_0^{v} \varepsilon_r(v') dv'.
\end{equation}
From this expression, we define capacitance as the ratio of a differential added charge to differential added voltage on the plates:
\begin{equation}
    C(v) \equiv \frac{\partial q}{\partial v} = \frac{ \varepsilon_0 \varepsilon_r(v) A}{d}.
    \label{eq:capacitance}
\end{equation}
Capacitance is proportional to permittivity, which in this case is voltage tunable.

To connect with nonlinear (and potentially quantum) optics, it is useful to derive expressions for the capacitor's stored energy $U_c$. This can be done either as a function of voltage to obtain $U_c(v)$, or as a function of charge to obtain $U_c(q)$:
\begin{align}
    U_c(v) &= \int_0^v q(v') dv', \\ 
    U_c(q) &= \int_0^q v(q') dq'. \label{eq:U_Q}
\end{align}
The first expression can be evaluated through $q(v) = \int_0^v C(v') dv'$ directly, while the second must be obtained by inverting $q(v)$ to obtain $v(q)$. This inversion can be done analytically with perturbative expansions, or numerically if the full behavior is needed.

\section{Circuit Hamiltonian}
\label{sec:circuit_hamiltonian}
The dynamics of an LC circuit may be expressed via Kirchoff's laws in terms of time-dependent voltages $v(t)$ and currents $I(t)$, or, the branch charge $\phi = \int_{-\infty}^{t} v(t') dt'$ across the inductor the and branch charge $q=\int_{-\infty}^{t} I(t') dt'$ across the capacitor. Expressed in terms of $\phi$ and $q$, the Hamiltonian is:
\begin{equation}
    H = \underbrace{\frac{\phi^2}{2L}}_\text{$U_l$} + \underbrace{\int_0^{q} v(q') dq'}_\text{$U_c$},
    \label{eq:H_appendix}
\end{equation}
where the capacitor's energy is modified from the usual quadratic to account for a nonlinear capacitance (i.e. a nonlinear charge-voltage relation). This Hamiltonian is comprised of both a linear component $H_\mathrm{harm}$ and nonlinear interaction terms $H_\mathrm{int}$, such that $H = H_\mathrm{harm} + H_\mathrm{int}$.

Introducing the offset charge $q_0$, induced by an external voltage bias $v_0$, shifts the potential. We can expand the resulting Hamiltonian via the Taylor expansion:
\begin{equation}
\begin{split}
    U_c(q_\delta) = U_c(q_0) + \frac{1}{2}U_c^{(2)}(q_0) q_\delta^2 + \sum_{n \ge 3} \frac{1}{n!} U_c^{(n)}(q_0) q_\delta^n,
    \label{eq:U_expansion}
\end{split}
\end{equation}
where $q_\delta = q - q_0$ is the differential charge around the equilibrium bias $q_0$, and $U_c^{(n)}$ is the $n^\mathrm{th}$ derivative of energy with respect to charge. By definition, the linear term in the expansion cancels out the contribution from the external bias. To lowest order the third- and fourth-order interaction terms control the nonlinear dynamics: $H_\mathrm{int} = \frac{1}{3!} U_c^{(3)}(q_0)q_\delta^3 + \frac{1}{4!} U_c^{(4)}(q_0)q_\delta^4 + \mathcal{O}(q_\delta^5)$.

It is often helpful to express Eq.~\ref{eq:H_appendix} in terms of the complex variables $a=\frac{1}{\sqrt{2\hbar}} \left(\frac{\phi}{\sqrt{z_0}} + i \sqrt{z_0} q_\delta \right)$ and $a^\dagger=\frac{1}{\sqrt{2\hbar}} \left(\frac{\phi}{\sqrt{z_0}} - i \sqrt{z_0} q_\delta \right)$ such that
\begin{align}
    \phi = \phi_\mathrm{zpf}(a + a^{\dagger}), \quad q_\delta = -iq_\mathrm{zpf}(a - a^{\dagger}), 
    \label{eq:operator_definitions}
\end{align}
where $\phi_\mathrm{zpf} = \sqrt{\hbar z_0/2}$ and $q_\mathrm{zpf} = \sqrt{\hbar/2 z_0}$ are the zero-point fluctuations of branch flux and charge, respectively \cite{vool:2017}. Although for now we treat $\phi$ and $q_\delta$ as classical fields, in the quantum mechanical picture these become operators with the commutation relations $[\hat{a},\hat{a}^\dagger]=1$ and $[\hat{\phi},\hat{q}_\delta]=i\hbar$. In general these variables depend on bias charge since the resonator impedance is $z_0(v)=\sqrt{L/C(v)}$. 


In order to analyze the underlying nonlinearity of our circuit, we first cast the total Hamiltonian in terms of the quadrature operators as follows:
\begin{align}
    H/\hbar = \omega_0a^\dagger a + ig_3\big(a - a^\dagger\big)^3 + g_4\big(a - a^\dagger\big)^4 + ...,
    \label{eq:Hamiltonian_Taylor_expansion}
\end{align}
where $\omega_0(v)=1/\sqrt{L C(v)}$ is the resonance frequency, and $g_3$ and $g_4$ are the third- and fourth-order nonlinearities. It follows from Eq.~\ref{eq:U_expansion} that
\begin{align}
    g_3 & = \frac{1}{3!} \frac{1}{\hbar} U^{(3)}_c q^3_\mathrm{zpf}, \\
    g_4 & = \frac{1}{4!} \frac{1}{\hbar} U^{(4)}_c q^4_\mathrm{zpf}.
\end{align}
The Taylor coefficients can be related to the external voltage bias $v_0$ keeping in mind that $\partial U_c/\partial q=v(q)$, and $\partial^2 U_c/\partial q^2=\partial v/\partial q=C^{-1}$. One can show
\begin{align}
    U^{(3)}_c(v_0) & = -C'(v_0) / C^3(v_0), \\
    U^{(4)}_c(v_0) & = \bigg(-C''(v_0) + 3\frac{C'(v_0)^2}{C(v_0)}\bigg)/C^4(v_0).
\end{align}
Thus we can directly express our nonlinearities in terms of our capacitance:
\begin{align}
    g_3(v_0) & = \frac{1}{\hbar} \frac{v^3_\mathrm{zpf} C'(v_0)}{3!} \label{eq:g3}, \\
    g_4(v_0) & = \frac{1}{\hbar} \frac{v^4_\mathrm{zpf}}{4!}\left(-C''(v_0) + 3\frac{C'(v_0)^2}{C(v_0)} \right)
    \label{eq:g4},
\end{align}
where $v_\mathrm{zpf} = q_\mathrm{zpf}/C(v_0)$ are the resonator's zero-point voltage fluctuations.

Three-wave mixing  (also sometimes referred to as squeezing in the parametric amplifier literature) arises when charge on the capacitor is modulated at twice the resonator's frequency. The typical approach to determine $\xi$ under the rotating-wave approximation (RWA) is to consider adding a modulation term to $q_\delta \to q_\delta + q_\mathrm{ac}\cos(2\omega_0 t+\theta)$ and expanding Eq.~\ref{eq:Hamiltonian_Taylor_expansion} keeping resonant terms \cite{eichler:2014,aumentado:2020}. This results in the Hamiltonian:
\begin{align}
    & H/\hbar = \omega_0 a^{\dagger}a + \frac{\xi}{2} a^{\dagger 2} + \frac{\xi^*}{2} a^{2} + \frac{K_\mathrm{eff}}{2} a^{\dagger 2}a^2 + \cdots, \label{eq:Hamiltonian_vs_q_appendix}
\end{align}
which appears in Eq.~\ref{eq:Hamiltonian_nonzero_bias} of the main text. Here the three-wave-mixing term $\xi$ is related to Eq.~\ref{eq:g3} as
\begin{align}
    \xi = 3 g_3 \left(\frac{v_\mathrm{ac}}{v_\mathrm{zpf}}\right) e^{-i\theta}
    \label{eq:3WM_strength_vs_v},
\end{align}
where $\theta$ is the phase of the pump tone and $v_\mathrm{ac} = q_\mathrm{ac}/C(v_0)$ is the ac modulation voltage. To lowest order the effective Kerr term $K_\mathrm{eff}$ is related to Eq.~\ref{eq:g4} as
\begin{align}
    K^\mathrm{RWA}_\mathrm{eff} = 12g_4.
    \label{eq:Kerr_eff_RWA}
\end{align}
One can also go further to second order in perturbation theory for Eq.~\ref{eq:Hamiltonian_Taylor_expansion} \cite{frattini:2021}, which results in 
\begin{align}
    K^\mathrm{2PT}_\mathrm{eff} = 12(g_4 - 5g^2_3/\omega_0).
    \label{eq:Kerr_eff_vs_v}
\end{align}
As shown in Fig. ~\ref{fig:gain}c of the main text, Eq.~\ref{eq:Kerr_eff_vs_v} therefore predicts an external bias where $K_\mathrm{eff}$ approaches zero.

\section{Design Considerations}
\label{sec:design_considerations}

In this section, we elucidate our circuit design considerations and attempt to separate out contributions from the PPC geometry and intrinsic material properties.

\begin{figure}[htb!] 
\begin{center}
\includegraphics[width=1.0\columnwidth]{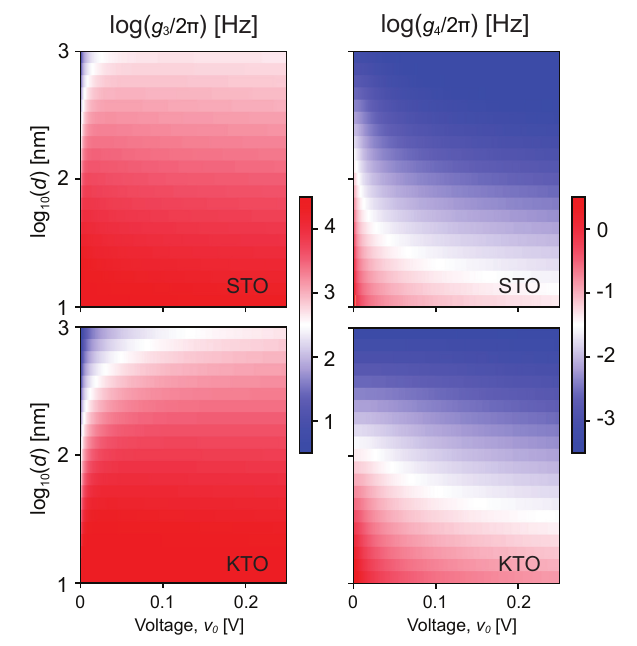}
\caption{Effect of varying the thin film thickness $d$ while keeping $A/d=(4~\units{\mu m})^2/(200~\units{nm})=8\times10^{-5}~\units{m}$ fixed for STO (top) and KTO (bottom). Thinner films have higher nonlinearities.
}
\label{fig:sweep_d}
\end{center}
\end{figure}

A critical design parameter is the STO or KTO film thickness $d$. In general, nonlinearities
are greater for smaller mode volumes where the field is
more concentrated. In Fig.~\ref{fig:sweep_d}, we analyze how the choice of $d$ impacts our circuit parameters while keeping the overall capacitance at zero external bias fixed (i.e. modifying $A$ and $d$ together such that $A/d$ remains constant). We observe that thinner capacitors result in higher nonlinearities, which translates to lower drive strengths required to obtain a target value of $\xi$, via Eq.~\ref{eq:3WM_strength_vs_v}. Understanding the trade-off between film thickness and internal loss will be key in optimizing performance.

\begin{figure}[htb!] 
\begin{center}
\includegraphics[width=1.0\columnwidth]{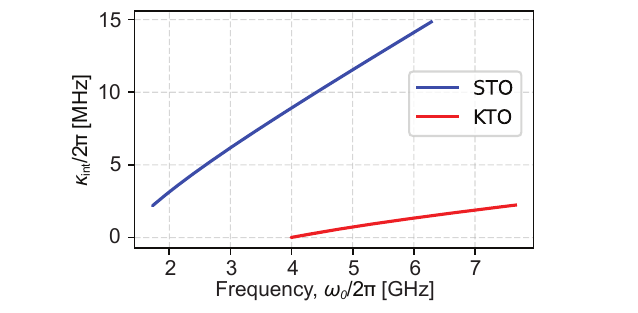}
\caption{
Internal loss rate $\kappa_\mathrm{int} = \omega_0 \times \mathrm{tan}(\delta)$ vs. resonator frequency, for both devices, where the loss tangent is determined from Landau-Ginzburg-Devonshire theory (see Fig.~\ref{fig:dielectric}c).
}
\label{fig:kappa_vs_freq}
\end{center}
\end{figure}

Operational bandwidth is another key design consideration. For a parametric amplifier based on a single resonant circuit this is set by the resonator's linewidth, $\kappa = \kappa_\mathrm{ext} + \kappa_\mathrm{int}$. Linewidth is a combination of external coupling rate $\kappa_\mathrm{ext}$ that may be near arbitrarily determined by choice of coupler, and, internal loss, which here we model to be dominated by the dielectric loss tangent $\kappa_\mathrm{int} = \omega_0 \tan(\delta)$. In Fig.~\ref{fig:kappa_vs_freq} we plot $\kappa_\mathrm{int}$ versus resonator frequency, for the dielectric model based on LGD theory as described in Appendix~\ref{sec:dielectric_properties}. In general, it is best to design a single-port parametric amplifier to be overcoupled, i.e. $\kappa_\mathrm{ext}\gg\kappa_\mathrm{int}$, because internal loss scales added noise by a factor of $(\kappa_\mathrm{int} + \kappa_\mathrm{ext})/\kappa_\mathrm{ext}$ \cite{eichler:2014}. For the parameters described here, chosen around an operation frequency at approximately $\units{GHz}$ frequencies, this suggests that an external coupling rate of 10's of MHz or greater will yield the best noise performance for both PANDA designs. 

The resonator's internal and external loss are related to its gain-bandwidth by solutions to Eq.~\ref{eq:gain_3WM}. To maximize linear gain it is useful to pump the circuit close to $2\xi/\kappa=1$, the limit of parametric instability, as plotted in Fig.~\ref{fig:gain}. The aforementioned constraints on $\xi$ thus determine that $\kappa_\mathrm{ext}$ cannot be arbitrarily increased without compromise to gain. As such, we predict in Fig.~\ref{fig:gain} that a gain of greater than 20 dB over order MHz of bandwidth is reasonable given design constraints. More bandwidth may require a more complicated circuit, paralleling work on broadband superconducting parametric amplifiers \cite{macklin:2015,malnou:2021,malnou:2022}.

\section{Nanofabrication}
\label{sec:nanofab}

\begin{figure*}[htb!] 
\begin{center}
\includegraphics[width=2.0\columnwidth]{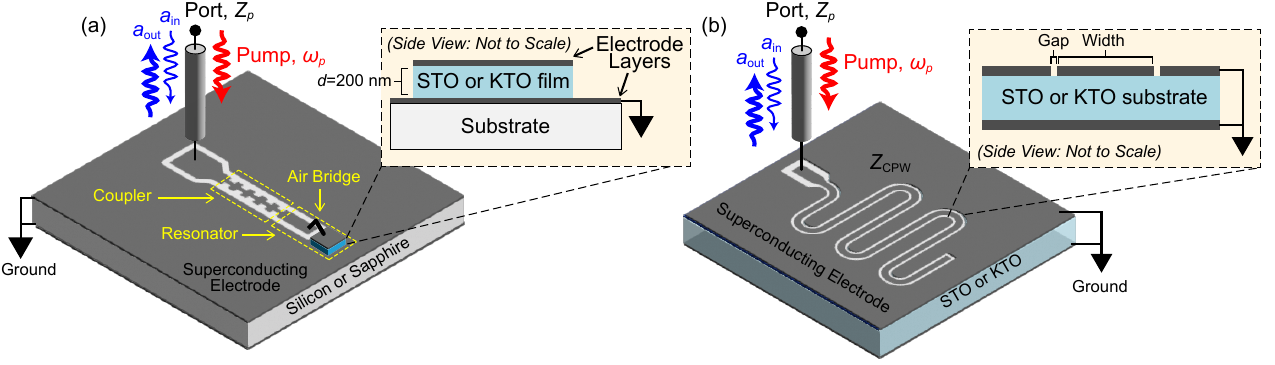}
\caption{
Example PANDA designs. (a) A parallel-plate capacitor is made from a thin layer STO or KTO sandwiched by bottom and top electrode layers. This structure sits on top of a conventional substrate, for example silicon or sapphire. Geometric inductance from the wiring layer combines with the parallel-plate capacitance to form a resonant circuit. The resonator is connected to a port using an inductive coupling element, e.g. a stepped-impedance filter as in Refs.~\cite{parker:2022,vaartjes:2024,frasca:2024}, which sets the external coupling rate $\kappa_\mathrm{ext}$. (b) Alternatively, a PANDA may be formed using a coplanar waveguide (CPW) resonator, like the device in Ref.~\cite{davidovikj:2017}, where a CPW of impedance $Z_\mathrm{CPW}$ is connected to a $Z_p=50 \units{\Omega}$ port. Because the CPW is fabricated on STO or KTO substrate, permittivity is high and $Z_\mathrm{CPW}$ can be very low compared to $50~\units{\Omega}$. This creates an impedance mismatch which sets the external coupling rate $\kappa_\mathrm{ext}$, forming a half-wavelength resonator. In contrast to a conventional CPW designed to have impedance of $50 \units{\Omega}$ , the gap (see inset) should be as small as possible compared to the width, to further reduce the impedance and to minimize the bias voltage needed to tune permittivity. In both (a) and (b), superconducting material (e.g. Nb) can be used to minimize loss and the conductor should be as thick as possible to minimize kinetic inductance. Charge on the center conductor can be controlled via a bias tee. 
}
\label{fig:layout}
\end{center}
\end{figure*}

Our proposed design is a lumped-element resonator formed by a parallel plate capacitor shunted by an external inductance, as in Ref.~\cite{zotova:2023}. A co-planar waveguide (CPW) structure \cite{davidovikj:2017}, shown for example in Fig.~\ref{fig:layout}, is able to leverage the same physics but with a weaker field confinement. In addition, the spatial dependence of the electric field within the CPW substrate complicates our theoretical treatment, because the relative orientation of the dc and rf electric fields will play a role in how much of the nonlinearity is activated. This effect is minimized for our lumped-element capacitor design.

To create the requisite parallel capacitor, thin films of STO or KTO can be directly grown on a variety of substrates, even including directly on a metallic superconducting layer \cite{schlom:2015,boikov:2000}. Likely, homoepitaxial growth will not be a viable route due to the large dielectric constant of the substrate. Depending on the substrate, epitaxial mismatch strain (i.e. with silicon substrates) will degrade the low-temperature dielectric properties. This can result in a lower tunability and alter the optimal operating temperature of the device.

Oxide films can also be grown on water-soluble buffer layers, allowing for layer transfer of the epitaxial thin films \cite{chen:2019}. Remote epitaxy on materials like graphene also can produce thin films \cite{yoon:2022}. Heterogeneous bonding and removal of the growth substrate (i.e. silicon) is an alternate route to transfer films. Finally, bulk thinning of crystals is another potential pathway to create low-strain, bulklike oxide thin films of a variety of substrates. Bonding methods can be interface-free and room temperature, compatible with existing devices on the substrate. Finally, to optimize performance, the tunability of the dielectric constant with field can be optimized with precise strain, doping, and isotope control. The exact coefficient of thermal expansion mismatch with the substrate will be critical.

For microwave devices, the oxide thin film needs to be laterally small and thin. The film would be transferred onto a conventional, low-loss substrates (e.g. sapphire or silicon) with a superconducting layer deposited on top. The resulting film on the superconductor-substrate stack would then be patterned into the desired capacitor size through dry or wet etching. The top surface would then be metalized with another superconducting layer. The bottom superconducting layer would connect to the circuit ground, while the top layer would be connected by a superconducting crossover to the substrate that leads to the on-chip meander inductor. As a result, small patterned coupons of thin-film oxides can be placed where they are desired on a standard superconducting device substrate and integrated seamlessly with existing circuitry. Flip-chip bonding techniques can also be used to place capacitors as desired on a microwave circuit \cite{engl:2019}.

While oxide thin films can be grown and transferred at the wafer scale, little is known about their low-temperature dielectric performance at high frequencies, requiring further study. Likely, thin-film properties will be degraded from the bulk. There may also be significant ``dead layer" effects, which limit the applicability of very thin films. Despite this, due to the proximity to the quantum phase transition, the dielectric constant will certainly still remain very large and tunable. Even with degraded performance from realistic materials, such devices may still constitute a competitive alternative to other cryogenic parametric-mixing elements.


\end{document}